# Turbulent Mixing, Viscosity, Diffusion, and Gravity in the Formation of Cosmological Structures: The Fluid Mechanics of Dark Matter

**C. H. Gibson**
Professor,
Departments of MAE and SIO, Center for Astrophysics and Space Sciences,
University of California at San Diego,
La Jolla, CA 92093-0411
e-mail, website: cgibson@ucsd.edu,
http://www-acs.ucsd.edu/~ir118

*Self-gravitational structure formation theory for astrophysics and cosmology is revised using nonlinear fluid mechanics. Gibson's 1996–2000 theory balances fluid mechanical forces with gravitational forces and density diffusion with gravitational diffusion at critical viscous, turbulent, magnetic, and diffusion length scales termed Schwarz scales. Condensation and fragmentation occur for scales exceeding the largest Schwarz scale rather than $L_J$, the length scale introduced by Jeans in his 1902 inviscid-linear-acoustic theory. The largest Schwarz scale is often larger or smaller than $L_J$. From the new theory, the inner-halo ($10^{21}$ m) dark-matter of galaxies comprises $\sim 10^5$ fossil-$L_J$-scale clumps of $10^{12}$ Earth-mass fossil-$L_{SV}$-scale planets called primordial fog particles (PFPs) condensed soon after the cooling transition from plasma to neutral gas, 300,000 years after the Big Bang, with PFPs tidally disrupted from their clumps forming the interstellar medium. PFPs explain Schild's 1996 ''rogue planets . . . likely to be the missing mass'' of a quasar lens-galaxy, inferred from twinkling frequencies of the quasar mirages, giving 30 million planets per star. The non-baryonic dark matter is super-diffusive and fragments at large $L_{SD}$ scales to form massive outer-galaxy-halos. In the beginning of structure formation 30,000 years after the Big Bang, with photon viscosity values $\nu$ of $5 \times 10^{26}\ m^2\ s^{-1}$, the viscous Schwarz scale matched the horizon scale ($L_{SV} \approx L_H < L_J$), giving $10^{46}$ kg proto-superclusters and finally $10^{42}$ kg proto-galaxies. Non-baryonic fluid diffusivities $D \sim 10^{28}\ m^2\ s^{-1}$ from galaxy-outer-halo ($L_{SD}$) scales ($10^{22}$ m) measured in a dense galaxy cluster by Tyson, J. A., and Fischer, P., 1995, ''Measurement of the Mass profile of Abell 1689,'' Ap. J., 446, pp. L55–L58, indicate non-baryonic dark matter particles must have small mass ($\sim 10^{-35}$ kg) to avoid detection.*
[S0098-2202(00)01504-2]

## Introduction

Information about the early universe has been flooding in with spectacular resolution from bigger and better telescopes on earth, on high altitude balloons, and in space, covering spectral bands previously unobservable. The 1989 COsmic Background Explorer (COBE) satellite and the other instruments reveal a color photograph of the universe as it existed 300,000 years after the hot Big Bang when the cooling opaque plasma formed a transparent H-He gas. Hubble Space Telescope (HST) images show evolved structures earlier than expected from standard linear cosmological models of Weinberg [1], Zel'dovich and Novikov [2], Silk [3,4], Kolb and Turner [5], Peebles [6], Padmanabhan [7], and Rees [8]. However, the most remarkable conclusion from all these observations of galaxies and galaxy clusters is that 99% or more of the matter is not in stars but is in unknown ''dark matter'' forms. What is this dark matter? From measured ratios of the light elements (H, D, He, Li) and nucleosynthesis theory of the Big Bang, only 4% or less of the total mass in a ''flat'' universe can be ordinary ''baryonic'' matter that interacts by the strong force, comprised of protons and neutrons in atoms and plasmas. The remaining 96% is truly exotic as an engineering fluid because it has the ghostly behavior of neutrinos that interact by the weak force, and may indeed be neutrinos. Most of the energy of the 1987 southern hemisphere supernova was radiated by a powerful neutrino blast wave that passed through the earth as though it wasn't there, producing only a handful of collisions in great pools of water and cleaning fluid buried under mountains in the northern hemisphere that serve as neutrino detectors. Several neutrino species have been identified and some of their mass differences have been measured; so that presently, neutrinos are the only known form of non-baryonic dark matter. Dozens of nonbaryonic particles have been proposed but none detected in a new discipline, termed astro-particle-physics, whose major goal is the solution of the non-baryonic dark matter problem. Gibson [9] suggests that many aspects of the dark matter paradox of astrophysics and cosmology may be resolved by a better job of fluids engineering. When was the first turbulence? What was the viscosity? Won't the weakly collisional non-baryonic dark matter have a large diffusivity that dominates and delays its gravitational instability, independent of its temperature and sound speed? Won't this render concepts like cold and hot dark matter obsolete? Present cosmology relies on Jeans's 1902 inviscid linear theory that reduces the problem of structure formation by gravity to one of gravitational acoustics.

In the following we review two theories of gravitational insta-

Contributed by the Fluids Engineering Division for publication in the JOURNAL OF FLUIDS ENGINEERING. Manuscript received by the Fluids Engineering Division June 15, 1999; revised manuscript received June 12, 2000. Associate Technical Editor: S. Banerjee.

bility. The linear acoustic Jeans theory is strongly modified by a nonlinear theory based on the mechanics of real fluids. Cosmological differences between the theories are discussed, and comparisons are made with observations. Finally, a summary and conclusions are provided.

## Problems With Jeans's Theory

Jeans's [10] theory of self-gravitational instability poorly describes this highly nonlinear phenomenon because its truncated momentum-mass equations and linear perturbation stability analysis exclude turbulence, turbulent mixing, viscous, Coriolis and magnetic forces, and molecular and gravitational diffusivity. In fluid mechanics it is well known that linear theories may give vast errors when applied to nonlinear processes. For example, neglect of the inertial-vortex forces in the Navier Stokes equations gives constant laminar flow velocity profiles that are independent of time and Reynolds number, contrary to observations that such flows always become turbulent when the Reynolds number exceeds a critical value. It is argued by Gibson [9,11–16] that the dark matter paradox is one of several cosmological misconceptions resulting from the application of Jeans's instability criterion to the development of structure by gravitational forces in the early universe.

Jeans's [10] theory predicts only acoustic instabilities, where sound waves of wavelength $\lambda$ require a time $\lambda/V_S$ to propagate a distance of one wavelength and gravitational response requires a free fall time of $(\rho G)^{-1/2}$, with $\rho$ the density, $G$ Newton's constant of gravitation, and $V_S$ the speed of sound. Sound waves provide density nuclei for condensation and fragmentation for wavelengths $\lambda \geqslant L_J$ by setting these two times equal to each other, so the Jeans criterion for gravitational instability of density perturbations on scale L is $L \geqslant L_J \equiv V_S/(\rho G)^{1/2}$. By the Jeans theory all density nuclei propagate with velocity $V_S$ as sound waves. Waves with $\lambda \leqslant L_J$ move away before gravity can act.

However, most density nuclei in natural fluids are nonacoustic, drifting with the local fluid speed $v \approx 0$ rather than $V_S$. Such density extrema generally result from turbulent scrambling of temperature and chemical species gradients, which produce density fluctuations $\delta\rho/\rho$ much larger than acoustic levels. The reference pressure fluctuation $\delta p$ for sound in air is $2 \times 10^{-5}$ kg/m s$^2$, corresponding to $\delta T/T$ of $6\times 10^{-11}$ and $\delta\rho/\rho$ of $1.4\times 10^{-10}$. Measurements by COBE of the cosmic microwave background (CMB) show $\delta T/T$ is $\sim 10^{-5}$. This small value proves the primordial plasma was not strongly turbulent, but it is large enough to make any acoustic interpretation (e.g., Hu [17]) highly questionable. Maximum local $\delta T/T \sim 10^{-4}$ values of 124 dB mapped by the BOOMERanG telescope (de Bernardis et al. [18]) nearly match those for the 125 dB sonic threshold of pain.

Because no local sources of sound existed in the hot plasma epoch, viscous damping was strong with short viscous attenuation lengths $V_s\lambda^2/\nu$, and because BOOMERanG $\delta T$ spectra show small or nonexistent harmonic ''sonic'' peaks it seems likely that all $\delta T/T$ fluctuations measured are nonacoustic. The observed spectral peak at subhorizon scales $L < L_H$ is more likely a fossil of the first gravitational structure, Gibson [9], nucleated by fossils of Big Bang ''turbulent'' $\delta T$ mixing in the quantum gravitational dynamics (QGD) epoch. QGD mixing is between a chaotic source of the space-time-energy (and $\delta T$) of the universe at the Planck temperature $T_P = (c^5 h/Gk^2)^{1/2} \approx 10^{32}$ $K$, Planck length scale $L_P = (hG/c^3)^{1/2} \approx 10^{-35}$ m and Planck time $t_P = (hG/c^5)^{1/2} \approx 10^{-43}$ s terminated by the strong force freeze-out temperature $10^{28}$ K at $10^{-35}$ s and $10^{-27}$ m, with inflation by a factor of $10^{49}$ to the fossil Planck scale $10^{14}$ m and fossil strong force horizon $L_H = 10^{22}$ m (see website http://www-acs.ucsd.edu/~ir118 for figures). The intermediate QGD ''turbulence'' $\delta T$ spectrum $k^2\phi_T = \beta\chi\varepsilon^{-1/3}k^{1/3}$ (not the Harrison-Zel'dovich spectrum $\phi_T \sim k^{-3}$) is fossilized by inflation (Guth [19]) stretching the fluctuations outside their scales of causal connection, where $\chi$ and $\varepsilon$ are dissipation rates of temperature and velocity variance that represent rates of entropy production at the beginning of inflation.

According to the turbulent mixing theory of Gibson [20], constant density surfaces move with the local fluid velocity except for their velocity $-\vec{r}D\nabla^2\rho/|\nabla\rho|$ with respect to the fluid due to molecular diffusion $D$, where $\vec{r}$ is a unit vector in the direction of the density gradient. The scale of the smallest density fluctuation is set by an equilibrium between this diffusion velocity D/L and the convection velocity $\gamma$L at distances L away from points of maximum and minimum density, giving the Batchelor scale $L_B \equiv (D/\gamma)^{1/2}$ independent of the ratio $Pr \equiv \nu/D$, where $\gamma$ is the local rate-of-strain and $\nu$ is the kinematic viscosity. This prediction has been confirmed by measurements in air, water and mercury for $0.02 \leqslant Pr \leqslant 700$ and by numerical simulations at smaller $Re \approx 2500$ for $10^{-2} \leqslant Pr \leqslant 1$, Gibson et al. [21]. Even if gravitational condensation of mass were to take place on a sound wave moving in a stationary fluid, it would immediately produce a nonacoustic density maximum from the conservation of momentum. Since the ambient condensing fluid is not moving, its momentum (zero) would immediately dominate the tiny momentum of the sound wave crest.

Gibson [9] shows that gravitational fragmentation at a nonacoustic density minimum or condensation on a non-acoustic density maximum is limited by viscous or turbulent forces at either the viscous Schwarz scale $L_{SV} \equiv (\gamma\nu/\rho G)^{1/2}$ or the turbulent Schwarz scale $L_{ST} \equiv \varepsilon^{1/2}/(\rho G)^{3/4}$, whichever is larger, where $\varepsilon$ is the viscous dissipation rate of the turbulence. For the superdiffusive non-baryonic dark matter that constitutes most of mass of the universe, the diffusive Schwarz scale $L_{SD} \equiv [D^2/\rho G]^{1/4}$ limits instability. The criterion for gravitational instability at scale $L$ is thus $L \geqslant L_{SX\,max} = \max[L_{SV}, L_{ST}, L_{SD}]$, where only viscous and turbulent forces are assumed to prevent instability in the early universe (magnetic forces are negligible) for the baryonic matter, and $L_{SD}$ sets the maximum scale for fragmentation of the non-baryonic matter ($\gamma < [\rho G]^{1/2}$). Because the universe is expanding, the largest scale structures form by fragmentation, which is assisted by the expansion, rather than condensation which is resisted.

$L_{SD} \equiv [D^2/\rho G]^{1/4}$ is derived by setting the diffusion velocity D/L equal to the gravitational velocity $L(\rho G)^{1/2}$. The diffusivity D of a gas is the particle collision length 1 times the particle velocity $v$. If $1 \geqslant L_H$ the particle is considered collisionless and more complex methods are required using the collisionless Boltzmann equation and general relativity theory. Density perturbations in collisionless species are subject to Landau damping, also termed collisionless phase mixing or free streaming, Kolb and Turner [[5] p. 351]. The free-streaming length $L_{FS}$ is about $10^{24}$ m for neutrinos assuming a neutrino mass of $10^{-35}$ kg corresponding to that required for a flat universe, giving an effective diffusivity of $3\times 10^{35}$ m$^2$ s$^{-1}$ from $L_{SD}$. Thus if neutrinos are the missing non-baryonic mass and collisionless, they are irrelevant to structure formation until $L_{FS} = L_H$ at $\sim 10^8$ years. From observations and $L_{SD}$ it appears that whatever the non-baryonic fluid may be, its diffusivity D is too small for its particles to be collisionless but too large for them to have the mass of any known particles besides neutrinos or they would have been observed.

In the early universe, the sound speed $V_S$ was large because of the high temperatures, and horizon scale Reynolds numbers Re $\approx c^2 t/\nu$ were small because the viscosity $\nu$ was large and $t$ was small. Therefore, $L_{SV}$ and $L_{ST}$ were both smaller than $L_J$, giving sub-Jeans-mass fragments in this period of time. From linear cosmology, no such fragmentation is possible with $t \leqslant 300{,}000$ years ($10^{13}$ s) in the plasma epoch following the Big Bang because $L_J > L_H \equiv ct$. No structures can form by causal processes on scales larger than $L_H$ because the speed of information transfer is limited by the speed of light c. Star formation is prevented by the Jeans criterion until the Jeans mass $M_J = (RT/\rho G)^{3/2}\rho$ decreases below a solar mass as the temperature of the universe decreases, but this requires $\sim 10^8$ years contrary to recent observations showing not only stars but galaxies and galaxy clusters existed at the earliest times observable; that is, at times $t < 10^9$ years with redshifts $z$ of 4 and larger. By the present nonlinear theory, viscous and turbulent forces permit fragmentations beginning at about 30,000 years ($t = 10^{12}$ s, $z = 4000$) when decreasing $L_{SV}$ values first match the increasing horizon scale $L_H$ with rate of strain $\gamma = 1/t$ and $\nu$ values more than $10^{26}\,m^2\,s^{-1}$, Gibson [11,12]. At this time the horizon mass $L_H{}^3\rho$ equaled the Schwarz viscous mass $M_{SV} = L_{SV}{}^3\rho$ at the observed supercluster mass of $10^{46}$ kg, Kolb and Turner [5], the largest structure in the universe. Density as a function of time can be computed from Einstein's equations of general relativity assuming a flat universe with kinetic energy always just matching gravitational potential energy, Weinberg ([1], Table 15.4). The horizon Reynolds number $c^2 t/\nu$ therefore was $\sim 150$, near the turbulent transition value.

This enormous $\nu = 10^{26}\,m^2\,s^{-1}$ can be explained as due to photon collisions with electrons of the plasma of H and He ions by Thomson scattering, with Thomson cross section $\sigma_T = 6.7 \times 10^{-29}\,m^2$. Fluctuations of plasma velocity are smoothed by the intense radiation since the ions remain closely coupled to the electrons by electric forces. We can estimate the kinematic viscosity $\nu \approx lc = 5 \times 10^{26}\,m^2\,s^{-1}$ using a collision length $l$ of $10^{18}$ m from $l = 1/\sigma_T n$, with number density $n$ of electrons about $10^{10}\,m^{-3}$ assuming the baryon (ordinary) matter density is $10^{-2}$ less than the critical density $\rho_C = 10^{-15}\,kg\,m^{-3}$ at the time ($\rho_C = 10^{-26}$ at present), from Weinberg [1]. Between 30,000 years and 300,000 years during the plasma epoch of the universe the temperature decreased from $10^5$ to 3000 K, decreasing the viscosity $\nu$ for the baryonic matter with its expansion gravitationally arrested at $\rho = 10^{-17}\,kg\,m^{-3}$, and decreasing the viscous Schwarz scale $L_{SV}$ of condensation due to decreases in both $\nu$ and $\gamma$. The final fragmentation mass by this scenario is about $10^{42}$ kg, the mass of a galaxy. As mentioned, the primordial plasma was not strongly turbulent from CMB observations of $\delta T/T \sim 10^{-5}$. If the flow were strongly turbulent, $\delta T/T$ values would be 3–4 orders of magnitude larger. Gravitational structure formation results in suppression of turbulence by buoyancy forces within the structures, Gibson [22].

Because the non-baryonic matter decouples from the fragmenting baryonic plasma by lack of collisional mechanisms while $L_{SD} \geq L_H$, it diffuses to fill the expanding proto-voids between the proto-superclusters, proto-clusters, and proto-galaxies developed during the plasma epoch, suppressing the gravitational driving force. The average density of galaxies of $10^{-21}\,kg\,m^{-3}$ today is $10^4$ less than the initial protogalactic baryonic density of $10^{-17}\,kg\,m^{-3}$ estimated in the present scenario, so galaxies never collapsed but expanded slowly and sometimes merged. The density $10^{-17}\,kg\,m^{-3}$ matches the density of globular star clusters, which is no coincidence since both $\rho$ and $\gamma$ at this time of first fragmentation should be preserved as hydrodynamic fossils, Gibson [14,15]. At some later time $\sim 10^8$ $y$ gravitational forces caused fragmentation of the non-baryonic matter at $L_{SD}$ scales to form halos of the evolving baryonic structures with galaxy to supercluster masses, as an effect rather than the cause.

Because the Jeans criterion does not permit baryonic matter to condense to form the observed structures, standard linear cosmology requires ''cold'' non-baryonic dark matter (CDM) to condense early in the plasma epoch, forming gravitational potential wells to guide the late collapse of the baryonic matter to form galaxies at $z \approx 5$ ($\sim 0.7 \times 10^9\,y$). This is accomplished by assuming the weakly interacting massive particles (WIMPs) have large masses, about $10^{-25}$ kg, giving small sound velocities and small Jeans CDM condensation masses in the galaxy mass range. Mixtures of such ''cold dark matter'' with less massive ''warm'' and ''hot'' dark matter particles are used to match observations of the actual universe structure. Such curve fitting is no longer required if the Jeans criterion is abandoned in favor of the recommended Schwarz length scale criteria.

## Theory of Gravitational Instability

Gravitational condensation for scales smaller than the horizon $L_H$ in the early universe can be described by the Navier Stokes equations of momentum conservation

$$\frac{\partial \vec{v}}{\partial t} = -\nabla B + \vec{v} \times \vec{\omega} + \overrightarrow{F_g} + \overrightarrow{F_v} + \overrightarrow{F_m} + \overrightarrow{F_{\text{etc.}}} \tag{1}$$

where $\vec{v}$ is the velocity, $B = p/\rho + v^2/2$ is the Bernoulli group, $\vec{v} \times \vec{\omega}$ is the inertial vortex force, $\vec{\omega}$ is the vorticity, $\overrightarrow{F_g}$ is the gravitational force, $\overrightarrow{F_v}$ is the viscous force, and the magnetic and other forces $\overrightarrow{F_m} + \overrightarrow{F_{\text{etc.}}}$ are assumed to be negligible. The gravitational force per unit mass $\overrightarrow{F_g} = -\nabla\phi$, where $\phi$ is the gravitational potential in the expression

$$\nabla^2 \phi = 4\pi\rho G \tag{2}$$

in a fluid of density $\rho$. The density conservation equation in the vicinity of a density maximum or minimum is

$$\frac{\partial \rho}{\partial t} + \vec{v} \cdot \nabla \rho = D_{\text{eff}} \nabla^2 \rho \tag{3}$$

where the effective diffusivity $D_{\text{eff}}$

$$D_{\text{eff}} = D - L^2/\tau_g \tag{4}$$

includes the molecular diffusivity $D$ of the gas and a negative gravitational term depending on the distance $L \geq L_{SX\,max}$ from the nonacoustic density nucleus and the gravitational free fall time $\tau_g = (\rho G)^{-1/2}$. Turbulence is driven by $\vec{v} \times \vec{\omega}$ forces.

For scales smaller than $L_H$ gravitational effects on space-time are described by Einstein's equations of general relativity

$$G_{ij} = R_{ij} - g_{ij} R = -(8\pi G/c^4) T_{ij} \tag{5}$$

where $R_{ij}$ is the Ricci tensor, $R$ is its trace, a term $\Lambda g_{ij}$ on the right has been set to zero (the cosmological constant $\Lambda$ introduced by Einstein and later dropped has recently been resurrected in attempts to reconcile observations with theory), G is Newton's gravitational constant, $T_{ij}$ is the energy-momentum tensor, $g_{ij}$ is the metric tensor, and indices $i$ and $j$ are 0, 1, 2, and 3. The Ricci tensor was developed to account for curvature problems of non-Euclidean geometry by Riemann and Christoffel and is formed by contracting the fourth-order curvature tensor so that Einstein's gravitational field tensor $G_{ij}$ on the left of (5) has only terms linear in second derivatives or quadratic in first derivatives of $g_{ij}$, Weinberg [[1] p. 153]. $G_{ij}$ was adapted from $R_{ij}$ to preserve Lorentz invariance and the equivalence of inertia and gravitation in mechanics and electromechanics. An expanding universe with critical density monotonically decelerates and is called flat by its mathematical analogy to zero curvature geometry. Classical solutions of the Einstein equations are given by standard cosmology texts such as Weinberg [1], Peebles [6], Kolb and Turner [5], Padmanabhan [7], and Coles [23].

The homogeneous-isotropic Robertson-Walker metric describes the universe after the Big Bang, where the cosmic scale factor $a(t) = R(t)/R(t_0)$ gives the time evolution of spatial scales in ''comoving'' coordinates $x' = x/a$ as the universe expands to the present time $t_0$. Variations in curvature of space can result in acausal changes of density for scales larger than $L_H$. Isocurvature fluctuations may not grow after inflationary expansion beyond the horizon and reenter the horizon at a later time with the same amplitude, Kolb and Turner [[5] p. 238]. Curvature fluctuations grow with the cosmological scale $R(t) \alpha t^{2/3}$ until they reenter the horizon. Strain rates diverge at zero time and in proper coordinates the horizon expands faster than $c$.

Jeans [10] considered the problem of gravitational condensation in a stagnant, inviscid gas with small perturbations of density, potential, pressure, and velocity so that the nonlinear term in (1) could be neglected along with all other terms except $\vec{F}_g$. He assumed that the pressure $p$ is a function only of the density $\rho$. Either the linear perturbation assumptions or this barotropic assumption are sufficient to reduce the problem to one of acoustics. Details of the Jeans derivation and its corrections from (5) are given in Kolb and Turner [[5] pp. 342–344] and in most other standard textbooks on cosmology, so they will not be repeated here. Diffusion terms are neglected in Eq. (3), and the adiabatic sound speed $V_S=(\partial p/\partial\rho)^{1/2}$ reflects an assumption that there are no variations in the equation of state. Cross differentiation with respect to space and time of the perturbed equations neglecting second order terms gives a wave equation for the density perturbation $\rho_1$

$$\frac{\partial^2\rho_1}{\partial t^2}-V_S^2\nabla^2\rho_1=4\pi G\rho_0\rho_1 \tag{6}$$

where $\rho_0$ is the unperturbed density. The solutions of (6) are of the form

$$\rho_1(\vec{r},t)=\delta(\vec{r},t)\rho_0=A\exp[-i\vec{k}\cdot\vec{r}+i\omega t]\rho_0 \tag{7}$$

which are sound waves of amplitude A for large $k\geqslant k_J$ which obey a dispersion relation

$$\omega^2=V_S^2k^2-4\pi G\rho_0 \tag{8}$$

where $k=|\vec{k}|$ and the critical wavenumber

$$k_J=(4\pi G\rho_0/V_S^2)^{1/2} \tag{9}$$

has been interpreted as the criterion for gravitational instability. All solutions of (6) with wavelength larger than $L_J$ are imaginary and are termed gravitationally unstable in linear cosmologies. Only such modes are considered to be eligible for condensation to form structure. Void formation is very badly modeled by linear cosmologies, and is not mentioned in standard treatments such as Kolb and Turner [5].

Consider the problem of gravitational instability for a nonacoustic density nucleus of diameter $L$ and mass $M'=\delta\rho L^3$, where $L_J>L>L_{SX\,max}$. For scales smaller than $L_J$ the pressure adjusts rapidly compared to the gravitational time $\tau_g\equiv(\rho G)^{-1/2}$. For scales larger than the largest Schwarz scale $L_{SX\,max}$ fluid mechanical forces and molecular diffusion are negligible compared to gravitational forces toward or away from the nucleus. Starting from rest, we see that the system is absolutely unstable to gravitational condensation or void formation, depending on whether $M'$ is positive or negative.

The radial velocity $v_r$ will be negative or positive depending on the sign of $M'$, and will increase linearly with time since the gravitational acceleration at radius $r$ from the center of the nucleus is constant with value $-M'G/r^2$. Thus

$$v_r=-M'Gt/r^2 \tag{10}$$

shows the mass flux

$$dM'/dt=-\rho v_r 4\pi r^2=M'\rho Gt4\pi \tag{11}$$

into or away from the nucleus is constant with radius. Integrating (11) gives two solutions

$$M'(t)=|M'(t_0)|\exp[\pm 2\pi\rho Gt^2]=|M'(t_0)|\exp[\pm 2\pi(t/\tau_g)^2] \tag{12}$$

where $M'(t_0)$ is the initial mass of the density nucleus. For condensation the only place where the density changes appreciably is at the core. We can define the core radius $r_c$ as

$$r_c=-v_r''t=M'Gt^2/r_c^2, \tag{13}$$

where $r_c$ is the distance from which core material has fallen in time t toward the core. The core mass change $M''$ is then

$$M''=\rho r_c^3=M'\rho Gt^2=M'(t_0)(t/\tau_g)^2\exp[2\pi(t/\tau_g)^2] \tag{14}$$

from (13) and (12).

For $M'<0$, the velocity of the rarefaction wave is limited by the sound speed $V_S$, but for $M'>0$ velocities become large for small $r$ according to (10). This may cause turbulence and inhibit condensation at times $t$ of order $\tau_g$, depending on $L_{SV}$ and $L_{ST}$.

The viscous Schwarz scale $L_{SV}$ is derived by setting the viscous force $F_V=\rho\nu\gamma L^2$ at scale $L$ equal to the gravitational force $F_g=G\rho L^3\rho L^3/L^2$, so

$$L_{SV}\equiv(\nu\gamma/\rho G)^{1/2} \tag{15}$$

where $\gamma$ is the rate of strain. Viscous forces overcome gravitational forces for scales smaller than $L_{SV}$. The turbulent Schwarz scale $L_{ST}$ is derived by setting the inertial vortex forces of turbulence $F_I=\rho V^2L^2$ equal to $F_g=G\rho^2L^4$, substituting the Kolmogorov expression $V=(\varepsilon L)^{1/3}$ for the velocity at scale $L$,

$$L_{ST}\equiv\varepsilon^{1/2}/(\rho G)^{3/4} \tag{16}$$

where $\varepsilon$ is the viscous dissipation rate of the turbulence. These two scales become equal when the inertial, viscous, and gravitational forces coincide. The gravitational inertial viscous scale

$$L_{GIV}=[\nu^2/\rho G]^{1/4} \tag{17}$$

corresponds to this equality, where $L_{GIV}=L_{SD}$ if $D=\nu$.

We can compare these expressions with the Jeans scale

$$L_J\equiv V_S/(\rho G)^{1/2}=[RT/\rho G]^{1/2}=[(p/\rho)/\rho G]^{1/2} \tag{18}$$

in terms of the temperature and pressure. The two forms for the sound velocity $V_S$ in (18) have led to the erroneous concepts of pressure support and thermal support, since by the Jeans criterion high temperature or pressure in a gas prevent the formation of structure. The length scale $L_{IC}\equiv[RT/\rho G]^{1/2}$ has the physical significance of an initial fragmentation scale in a uniform gas based on the ideal gas law $p=\rho RT$, where decreases in density are matched by rapid decreases in pressure so that the temperature remains constant on scales less than $L_J$. Thus cooling occurs for $L\geqslant L_{IC}$ when the finite speed of sound limits the pressure adjustment in such voids, so that fragmentation is accelerated by radiative heat transfer from the warmer surroundings. The length scale $L_{HS}\equiv[(p/\rho)/\rho G]^{1/2}$ is a hydrostatic scale that arises if an isolated blob of gas approaches hydrostatic equilibrium, with zero pressure outside. Neither $L_{IC}$ nor $L_{HS}$ have any physical connection to Jeans's theory. $L_{IC}$ and $L_{SV}$ are the initial gravitational fragmentation scales of the primordial gas to form PGCs and PFPs. $L_{HS}$ appears at the final stages of primordial-fog-particle formation as their size, but as an effect of the formation not the cause.

## Cosmology

The conditions of the primordial gas emerging from the plasma epoch are well specified. The composition is 75% hydrogen-1 and 25% helium-4 by mass, at a temperature of 3000 K. This gives a dynamical viscosity $\mu$ of $2.4\times10^{-5}\ \mathrm{kg\,m^{-1}\,s^{-1}}$ by extrapolation of the mass averaged $\mu$ values for the components to 3000 K, so the kinematic viscosity $\nu=\mu/\rho$ depends on the density $\rho$ assumed. The most likely $\rho$ for the baryonic gas is $10^{-17}\ \mathrm{kg\,m^{-3}}$, the fossil-density-turbulence value at the time $10^{12}$ s of first structure formation, with fossil-vorticity-turbulence $\gamma\approx10^{-12}\ \mathrm{s^{-1}}$. We find $L_{SV}=(\nu\gamma/\rho G)^{1/2}=6\times10^{13}$ m and $M_{SV}=L_{SV}{}^3\rho=2\times10^{24}$ kg as the most likely mass of the first gravitationally condensing objects of the universe, termed primordial fog particles or PFPs. With this density the gravitational time $\tau_g=(\rho G)^{-1/2}$ is $4\times10^{13}$ s, or 1.3 million years. However, the time required for the voids to isolate the individual PFPs should be much less since the speed of void boundaries represent rarefaction waves, and may thus approach the sound speed $V_S = 3\times10^3\ \mathrm{m\,s^{-1}}$, giving an isolation time $L_{SV}/V_S = 2\times10^{10}$ s, or 700 years. The viscous dissipation rate $\varepsilon = \nu\gamma^2 = 2\times10^{-12}\ \mathrm{m^2\,s^{-3}}$. The range of estimated PFP masses for various densities and turbulence levels $10^{24}$ to $10^{26}$ kg includes the Earth-mass of $6.0\times10^{24}$. Kolmogorov and Batchelor scales $L_K = L_B = 1.4\times10^{12}$ m.

Thus the entire baryonic universe of hydrogen and helium gas rapidly turned to PGC clumps of fog as the cooling plasma universe neutralized, with resulting primordial fog particle masses near that of the earth, separated by distances about $10^{14}$ m ($10^3$ AU). These PFPs constitute the basic materials of construction for everything else. Those that have failed to accrete to star mass, and this should be about 97%, constitute the baryonic dark matter. The mass of the inner halos of galaxies should be dominated by the mass of such PFPs, since the non-baryonic component diffuses to $L_{SD}$ scales that are much larger.

## Observations

Quasars are the most luminous objects in the sky. They are generally thought to represent black holes in cores of cannibal galaxies at an early stage of their formation when they were ingesting other galaxies, one or two billion years after the Big Bang. Quasar microlensing occurs when a galaxy is precisely on our line of sight to the quasar, so that it acts as a gravitational lens. The quasar image is split into two or more mirage-like images which twinkle at frequencies determined by the mass of the objects making up the lens galaxy. Schild [24] reports the results of a 15 year study of the brightness fluctuations of the two images of the QSO Q0957+561 A,B gravitational lens, amounting to over 1000 nights of observations. The time delay of 1.1 years was determined to subtract out any intrinsic quasar variability. The dominant microlensing mass was shown by frequency analysis to be $6.3\times10^{24}$ kg, close to the primordial fog particle mass estimated above and by Gibson [9]. Three observatories have since independently reported the same time delay and microlensing signals for this lensed quasar. Thus it is an observational fact that the mass of at least one galaxy is dominated by planetary mass objects, with $3\times10^7$ planets per star. Star-microlensing searches for planetary mass MACHOs (massive compact halo objects) have failed because PFPs are sequestered in PGC clumps, and have highly intermittent lognormal particle density distributions as a result of their nonlinear self-similar gravitational accretion cascades to form larger objects and ultimately stars, Gibson and Schild [16].

Planetary nebula (PN) appear when ordinary stars are in a hot dying stage on their way to becoming white dwarfs. Strong stellar winds and intense radiation from the central star should cause ambient PFPs to reevaporate and reveal themselves. Hubble Space Telescope observations of the nearest planetary nebula Helix (NGC 7293), by O'Dell and Handron [26], show a halo of $>6500\sim10^{25}$ kg "cometary knots" with tails pointing away from hot central star like "comets brought out of cold storage." Thousands of PFP-like "particles" also appear in HST photographs (PRC97-29, 9/18/97) of the recurring Nova T Pyxidis by M. Shara, R. Williams, and R. Gilmozzi, and as radial "comets" in recent HST photographs of the Eskimo PN (NGC 2392, A. Fructer et al., PRC00-07, 1/24/00).

Tyson and Fischer [25] report the first mass profile of a dense galaxy cluster Abel 1689 from tomographic inversion of 6000 gravitational arcs of 4000 background galaxies. The mass of the cluster is $10^{45}$ kg, with density $5\times10^{-21}\ \mathrm{kg\,m^{-3}}$. From the reported mass contours the cluster halo thickness is about $6\times10^{21}$ m. Setting this size equal to $L_{SD} = [D^2/\rho G]^{1/4}$ gives a diffusivity $D = 2\times10^{28}\ \mathrm{m^2\,s^{-1}}$, more than a trillion times larger than that of any baryonic gas component. A virial particle velocity $v = (GM/r)^{1/2} = 3.3\times10^6\ \mathrm{m\,s^{-1}}$ with mean collision distance $l = D/v = 6\times10^{21}\ m = m_p/\rho\sigma$ gives a collision cross section $\sigma = m_p(GM/r)^{1/2}/\rho D = 10^{-37}\ \mathrm{m^2}$ taking a particle mass $m_p = 10^{-35}$ kg corresponding to the neutrino mass required to produce a flat universe. If the particles are nearly relativistic, l is $7\times10^{19}$ m and $\sigma = 10^{-35}\ \mathrm{m^2}$.

## Conclusions

The Jeans length $L_J$ determines the formation of Jeans-mass-PGCs (Proto-Globular-star-Clusters), but it overestimates the minimum mass of baryonic fragmentation by 2–5 orders of magnitude during the plasma epoch when proto-supercluster to proto-galaxy objects were formed, and by 12 orders of magnitude in the hot gas epoch for PFPs (Primordial-Fog-Particles). In the cold, dense, turbulent molecular clouds of galactic disks where modern stars condense, the Jeans mass is generally smaller than the turbulent Schwarz mass by an order of magnitude, and is therefore irrelevant. The Gibson [11–16] hydro-gravitational criteria are recommended instead of Jeans's; that is, $L \geq L_{SX\,max} = \max[L_{SV}, L_{ST}, L_{SD}]$, where structure formation occurs at scales $L$ larger than the largest Schwarz scale.

According to the new theory, gravitational structure formation in the universe began in the plasma epoch at a time about 30,000 years after the Big Bang with the formation of proto-supercluster-voids and proto-superclusters in the baryonic component, triggered by inflated fossil "turbulence" density fluctuations from the QGD epoch ($t<10^{-35}$ s). The fragmentation mass decreased to that of a proto-galaxy by the time of plasma neutralization at 300,000 years. Immediately after atoms formed, the baryonic fragmentation mass decreased to that of a small planet and the universe of neutral primordial hydrogen and helium gas turned to fog within $L_J$ scale PGCs. Some of these primordial fog particles have aggregated to form stars and everything else, but most are now frozen as rogue planets and sequestered in clumps within PGC clumps as the dominant form of dark matter within $10^{21}$ m (30 kpc) galaxy-inner-halos. Many PFPs have probably been disrupted from their PGCs by tidal forces to form the dominant interstellar mass component of galaxy disks and galaxy cores. The non-baryonic part of the universe does not drive the formation of baryonic structures even though it is more massive, but is instead driven by them. It is highly diffusive from its small collision cross section $\sigma$, and diffuses to large $L_{SD}$ scales near $10^{22}$ m to form the outer halos of isolated galaxies and galaxy cluster halos in response to these large baryonic structures. Indicated $\sigma$ values near $10^{-36}\ \mathrm{m^2}$ are reasonable for small particles like neutrinos, but $\sigma = m_p(GM/r)^{1/2}/\rho D \approx 10^{-26}\ \mathrm{m^2}$ indicated for more massive non-baryonic candidates like $10^{-25}$ kg neutralinos are much larger than theoretical $\sigma\approx10^{-46}$ values or recent $\sigma\leq10^{-42}\ \mathrm{m^2}$ values excluded by observations using sensitive WIMP detectors (Dark Matter 2000 Conference, Marina del Rey, February 2000).

## Nomenclature

AU = astronomical unit (solar distance), $1.4960\times10^{11}$ m
$a(t)$ = cosmological scale factor as a function of time t
$c$ = speed of light, $2.9979\times10^8\ \mathrm{m\,s^{-1}}$
$D$ = molecular diffusivity of density, $\mathrm{m^2\,s^{-1}}$
$\varepsilon$ = viscous dissipation rate, $\mathrm{m^2\,s^{-3}}$
G = Newton's gravitational constant, $6.7\times10^{-11}\ \mathrm{m^3\,kg^{-1}\,s^{-2}}$
$\gamma$ = rate of strain, $\mathrm{s^{-1}}$
$k_B$ = Boltzmann's constant, $1.38\times10^{-23}\ \mathrm{J\,K^{-1}}$
$h$ = Planck's constant, $2\pi\ 1.05\times10^{-34}\ \mathrm{kg\,m^2\,s^{-1}}$
$l$ = collision length, m
ly = light year, $9.461\times10^{15}$ m
$L_{GIV}$ = gravitational-inertial-viscous scale, $[\nu^2/\rho G]^{1/4}$
$L_{SV}$ = viscous Schwarz scale, $(\nu\gamma/\rho G)^{1/2}$
$L_{ST}$ = turbulent Schwarz scale, $\varepsilon^{1/2}/(\rho G)^{3/4}$
$L_{SD}$ = diffusive Schwarz scale, $(D^2/\rho G)^{1/4}$
$L_H$ = Hubble or horizon scale of causal connection, ct
$L_J$ = Jeans scale, $V_S/(\rho G)^{1/2}$

$\lambda$ = wavelength, m
$m_p$ = proton mass, $1.661\times10^{-27}$ kg
$M_{\text{sun}}$ = solar mass, $1.99\times10^{30}$ kg
$M_{\text{earth}}$ = earth mass, $5.977\times10^{24}$ kg
$\upsilon$ = kinematic viscosity, $m^2\,s^{-1}$
pc = pars, $3.0856\times10^{16}$ m
$R$ = gas constant, $m^2\,s^{-2}\,K^{-1}$
$R(t)$ = cosmological scale, $a(t)=R(t)/R(t_0)$
$\rho$ = density, $kg\,m^{-3}$
$\rho_C$ = critical density, $10^{-26}\,kg\,m^{-3}$ at present for flat universe
$\sigma$ = collision cross section, $m^2$
$\sigma_T$ = Thomson cross section, $6.6524\times10^{-29}\,m^2$
$t$ = time since Big Bang
$t_0$ = present time, $4.6\times10^{17}$ s
$T$ = temperature, K
$V_S$ = sound speed, $m\,s^{-1}$
$z$ = redshift$=\lambda/\lambda_0-1$